\documentstyle[aps,prl,psfig,multicol,epsf]{revtex}
\draft
\newcommand{\be}{\begin{equation}}
\newcommand{\ee}{\end{equation}}
\newcommand{\bea}{\begin{eqnarray}}
\newcommand{\eea}{\end{eqnarray}}
\def\jp{J/\psi}
\def\p{\psi}

\def\ba{\begin{eqnarray}}
\def\ea{\end{eqnarray}}

\begin{document}
\title{Charm-sea Contribution to High-$p_T$ $\psi$ 
Production at the Fermilab Tevatron
\\[2mm]} 
\author{Cong-Feng Qiao\footnote{JSPS Research Fellow. 
E-mail: qiao@theo.phys.sci.hiroshima-u.ac.jp}}
\address{Department of Physics, Faculty of Science,\\
Hiroshima University, Higashi-Hiroshima 739-8526, Japan}
\maketitle

\begin{abstract}
The direct production of $J/\psi(\psi')$ at large 
transverse momentum, $p_T \gg M_{J/\psi}$, at the Fermilab 
Tevatron is revisited. It is found that the sea-quark 
initiated processes dominate in the high-$p_T$ region 
within the framework of color-singlet model, which is 
not widely realized. We think this finding is 
enlightening for further investigation on the 
charmonium production mechanism.

\end{abstract}
\begin{multicols}{2}

Quarkonium production and decays have long been taken 
as an ideal means to investigate the nature of Quantum 
Chromodynamics(QCD) and other phenomena. Due to the 
approximately non-relativistic nature, the description 
of heavy quark and antiquark system stands as one of the 
simplest applications of QCD. The rich spectrum of its 
radial and orbital excitations provides a suitable play 
ground for testing QCD based models. The heavy, but not 
very heavy, quark mass enables one to get knowledge of 
both perturbative and nonperturbative QCD via investigating 
quarkonium production and decays. The clean signals 
of quarkonium leptonic decays render the experiment detection 
with a high precision, and therefore, quarkonium may play an 
unique role in the study of other phenomena as well, 
e.g. in detecting the parton distribution, the QGP signal,
and even new physics. However, only with a theory 
which can precisely describe heavy quarkonium production 
and decays, may these advantages come true.

During the past decade, intrigued by the discovery of 
$\jp(\p')$ surplus production at high $p_T$ at the 
Fermilab Tevatron \cite{cdf1,cdf3}, our understanding 
on the natures of quarkonium production and decays has 
experienced dramatic changes. 

Conventionally, the so-called color-singlet model (CSM) 
was widely employed in the study of heavy quarkonium
production and decays \cite{t.a.degrand}. In CSM, 
it is assumed that the $Q\bar{Q}$ pair produced in a high 
energy collision will bind to form a given quarkonium state
only if the $Q\bar{Q}$ pair is created in color-singlet 
state with the same quantum numbers as the produced bound 
states; as well, in the quarkonium decays the annihilating 
$Q\bar{Q}$ pair will be in short distance and singlet with 
the same quantum numbers as its parent bound states.
It is assumed in CSM that the production amplitudes 
can be factorized into short distance and long distance 
parts. The short distance sector is perturbative QCD
applicable, while all the long distance nonperturbative 
effects are attributed to a single parameter, the wave 
function. That is, e.g., 
\ba
d\sigma(\psi_n + {\sc{X}}) = 
d\sigma (c\bar{c_1}(^3S_1) + {\sc{X}}) |R_{\psi_n}(0)|^2\;.
\ea
The wavefunctions can be either determined phenomenologically 
through experiment measurement of quarkonium leptonic decay rates, 
like
\ba
\Gamma(\psi_n \rightarrow l\bar{l})\approx \frac{4
\alpha^2} {9 m_Q^2} |R_{\psi_n}(0)|^2 \;,
\ea
or calculated from potential models.

CSM provides a prescription for calculations of not only 
the inclusive production rate of quarkonium states, but also 
their inclusive decay rates into light hadrons, leptons, 
and photons. Based on it many investigations had been carried
out in past more than two decades, and at least qualitative 
description of quarkonium production and decays were 
achieved. Nevertheless, color-singlet factorization is only 
an ad hoc hypothesis. There are no general arguments to 
guarantee such a naive model may still work up to higher 
order radiative corrections. In fact people got know for
quite long time it does not. Furthermore, the attempt 
to incorporate relativistic corrections also met difficulties 
\cite{barbieri}. Although this kind of shortcomings of CSM
had been known for many years, and some disagreements between
theoretical predictions and experiment data existed, there was 
no big breakthrough in theory until the CDF group 
released \cite{cdf1} the data on large-$p_T$ $J/\psi$
production collected in the 1992-1993 run. The new data, 
which benefited from the advanced technology of vertex detector were 
free of the large background from $B$ decays, put the CSM in an 
awkward situation, as the data differ very much from the leading 
order(LO) CSM predictions in both normalization and $p_T$ scaling. 

In 1993, Braaten and Yuan \cite{braaten1} noticed that at 
sufficiently high $p_T$ the dominant charmonium production 
mechanism is the production of a parton with large transverse 
momentum followed by its fragmenting into a charmonium state. 
With including the contribution by fragmentation mechanism, 
the prompt $J/\psi$ data can be explained within a small 
amount of error \cite{cacciari,braaten2}, where the charm 
quark splitting into $\chi_c$ states and then feeding down 
to $J/\psi$ contributes overwhelmingly. Nevertheless, one 
can merely get a similar $p_T$ asymptotic behavior for $\psi'$ 
production with the same scenario, and large discrepancy in 
normalization remained as well. This phenomenon was referred 
as the so-called "$\psi'$-surplus" production or "-anomaly". 

In 1997, a measurement of direct $J/\psi$ production 
exposed \cite{cdf3}, in which the higher excited states 
feeddown were stripped off. To one's surprise
the new experiment result excesses the CSM 
prediction by a factor of $\sim$ 30, the same as 
in $\psi'$ production. Nowadays, the former 
"$\psi'$-surplus" problem turns to be the generic 
"$\psi$-surplus" problem terminologically.

A general factorization formalism \cite{nrqcd} developed
from the non-relativistic QCD(NRQCD) \cite{caswell}, 
which describes the inclusive heavy quarkonium production 
and decays, were established by Bodwin, Braaten and Lepage(BBL). 
NRQCD is formulated from first principles and the BBL approach 
allows relativistic and radiative corrections to be performed 
safely to any desired order. One of the striking 
advancements of the new development from CSM is that within 
the BBL framework the intermediate $Q\bar{Q}$ state, which 
subsequently evolves into quarkonium states nonperturbatively, 
can be in both color-singlet and -octet configurations. At 
first order in $v$, the relative velocity of heavy quark, 
BBL and CSM are coincident in describing the S-wave quarkonium 
production.

Based on the BBL formalism, Braaten and Fleming suggested to 
solve the $\psi'$ surplus production puzzle via color-octet 
mechanism(COM) \cite{fleming}. They proposed that the 
dominant $\psi'$ production at high $p_T$ is through the 
fragmentation of a gluon into a $c\bar{c}$ pair in 
color-octet configuration, which will evolve into 
$\psi'$ non-perturbatively. Indeed they gave a well-fitted 
curve to the data and from which the non-perturbative matrix 
element $<{\cal{O}}_8^{\psi'}(^3S_1)>$ was extracted with
a magnitude being consistent with the estimation from
NRQCD "velocity scaling rules". After their pioneer work, 
hundreds of investigations have been performed in order 
to find either the signatures of color-octet states or 
any implication of the new proposal to other phenomena 
\cite{rothstein}.

The present situation is that on one hand the COM 
stands as the most plausible approach, up to now,
in explaining the $\jp(\psi')$ production "anomaly"; 
on the other hand, this scenario encounters some difficulties 
in confronting with other phenomena \cite{rothstein}. 
The most striking crisis is the absence of high-$p_T$ 
transversely polarized $J/\psi$ and $\psi'$ at the Tevatron 
in the first measurement from CDF \cite{cdf4}. According to 
NRQCD spin-symmetry, and the prescription that the dominant 
charmonium production mechanism at high $p_T$ is of a gluon 
splitting into a color-octet $^3S_1$ charm quark pair, such 
polarized states should appear \cite{wise}. Therefore, to what 
degree the COM plays the role in quarkonium production is 
still an open question to my understanding. To find 
distinctive color-octet signatures and to eliminate the 
large errors remaining in different fits for corresponding
matrix elements are currently urgent tasks in this research 
realm, for both theory and experiment.

In order to overcome the difficulties COM met, 
people tried to attribute large amount of high-$p_T$ 
events to intrinsic transverse momentum of the 
interacting partons, suppose that the large uncertainties 
existed in the $k_t$-factorization are manageable and the 
$k_t$ would still manifest itself in not very small-$x$ 
\cite{hagler}. To be noted that in the $k_t$-factorization
formalism, the analyses suggest that the direct $J/\psi$
production is still dominated by color-octet contributions, 
but from $^1S_0^{(8)}$ and $^3P_J^{(8)}$, up to large 
transverse momenta of the order $p_T \le 20$ GeV.

Now that the difficulties for direct $J/\psi$ 
and its radial excitation $\psi'$ production at the 
Tevatron are the same within a small amount of error, as 
aforementioned, one may reasonably infer that the origins 
accounting for the large discrepancies between experimental
data and the color-singlet description for both states 
would be the same. On this premise our investigation 
in this work will be restricted to $J/\psi$ for simplicity. 
The results and conclusions are applicable to $\psi'$.

We notice that within the framework of collinear 
factorization, quarkonium production processes initiated 
by the sea-quark interactions have been paid less 
attention in previous investigations in color-singlet 
prescription. In Ref. \cite{kniehl}
the sea initiated processes were considered, indirectly, 
for the large-$p_T$ $J/\psi(\psi')$ production, where
the relative importance of valence and sea parton
interacting processes was not distinguished. The 
leading order(LO) charm-sea interacting process was 
investigated in ref. \cite{saleev}, and found that it 
contributes negligiblely to the large-$p_T$ $\psi$
production as expected, since CDF 
data indicates that the $d\hat{\sigma}/dp_T^2$ scaling 
favors fragmentation process behaving like $1/p_T^4$ 
which happens beyond the LO. 
Therefore, in order to estimate the sea quark contributions  
we need to consider the possible processes beyond the LO. 
It is well-known that the source of the charm-sea
distribution can be traced back to higher order 
gluon-gluon processes. And, because of the large 
logarithmic term remaining in the NLO result of
gluon-gluon to charm pair \cite{nlo}, the evolution 
effects of summing up all the large logs to any order tend to 
be important for the processes we are interested in here. 
Indeed, following evaluation shows that the naive NLO gluon-gluon 
to charm pair partonic process, which was considered in the 
calculations of \cite{cacciari,braaten2}, can not simply 
substitutes the charm-sea induced processes of our concern. 

On the other hand, although superficially the sea-quark 
interacting processes seems to be negligible, since the sea 
distribution probabilities are pretty small comparing to those 
of valence quarks and gluons. Due to being in high energy and 
at high, but not very high, $p_T$ region, the $\hat{s}/\hat{t}$, 
the ratio of Mandelstam variables of the total center-of-mass 
energy squared in $s$ and $t$-channels, 
kinematically suppresses the valence-quark and gluon initiated 
processes relative to the sea interacting ones. And in the meantime 
high energy and large $p_T$ enhance the sea quark densities 
inside the incident hadrons. To see this picture more clearly, 
let us have a close look at the fragmentation prescription 
for quarkonium production. Generally, quarkonium fragmentation 
production, $A\; +\; B \rightarrow \psi\; + \; {\sc X}$, can be 
expressed as
\bea
d\sigma (A\; &+&\; B \rightarrow H(p_T)\; +\; {\sc X}) =
\sum_{a,b,c} \int_0^1 dx_a f_{a/A}(x_a) \nonumber \\ 
&\times& \int_0^1 dx_b f_{b/B}(x_b) \int_0^1 dz
d\hat{\sigma}(a + b \rightarrow c(p_T/z) + {\sc X})
\nonumber\\
&\times& D_{c \rightarrow H}(z,\; \mu)\;,
\label{eq0}
\eea
where $c$ is the fragmenting parton, either a gluon or
a (anit)charm quark, and the sum runs over all possible
partons. $D(z, \mu)$ is the fragmentation
function and $z$ is the momentum fraction of the
the fragmenting parton carried by quarkonium state.
The evolution of the fragmentation function 
$D_{c \rightarrow H}(z,\; \mu)$ with scale $\mu$ 
in Eq.(\ref{eq0}) is accomplished by the utilization of 
Alterelli-Parisi(AP) equations
\bea
\mu \frac{\partial}{\partial \mu} D_{i\rightarrow \psi}
(z,\; \mu) = \sum_j \int_z^1 \frac{dy}{y} P_{ij}(z/y,\mu)
D_{j\rightarrow \psi}(y,\; \mu)\;,
\label{ev}
\eea
where the $P_{ij}$ are the splitting functions of a parton
$j$ into a parton $i$.

To show the importance of the sea quark interacting processes 
in the fragmentation approach (\ref{eq0}), we do a simple 
comparison, for example, of the hard-scattering processes  
\bea
g + g \rightarrow C + \bar{C}
\label{eq1}
\eea
with
\bea
g + C(\bar{C}) \rightarrow g + C (\bar{C})\;,
\label{eq2}
\eea
where the $C$ and $\bar{C}$ stand for charm
and anticharm quarks, which are produced slightly 
off-shell and in high energy with large $p_T$. In LO
and massless limit, the differential cross sections 
for processes (\ref{eq1}) and (\ref{eq2}) are:
\bea
\frac{d\sigma_5}{d\hat{t}} (\hat{s},\hat{t}) =
\frac{\pi \alpha_s^2}{\hat{s}^2} \left\{
\frac{1}{6}\left(\frac{\hat{u}^2 + \hat{t}^2}
{\hat{u}\hat{t}}\right) - \frac{3}{8}\left(
\frac{\hat{u}^2 + \hat{t}^2}{\hat{s}^2}\right)\right\}
\label{eq3}
\eea
and
\bea
\frac{d\sigma_6}{d\hat{t}} (\hat{s},\hat{t}) =
\frac{\pi \alpha_s^2}{\hat{s}^2} \left\{
\left(\frac{\hat{u}^2 + \hat{s}^2}
{\hat{t}^2}\right) - \frac{4}{9}\left(
\frac{\hat{u}^2 + \hat{s}^2}{\hat{s}\hat{u}}\right)
\right\}\;,
\label{eq4}
\eea
respectively. Since we are interested in large-$p_T$ 
$J/\psi(\psi')$ hadroproduction at high energy, obviously
in a certain scope of phase space, the process (\ref{eq1})
is suppressed with respect to process (\ref{eq2}) by the 
factor of $\hat{s}/\hat{t}$. However, this is just a 
schematic argument. To be more strict, we need to convolute 
the hard scattering cross section with parton 
distribution and fragmentation functions. Without 
losing qualitative correctness, for simplicity 
we do a comparison for the subsets 
$g + g \rightarrow C+\bar{C}$
and 
$g + C(\bar{C}) \rightarrow g + \;C(\bar{C})$ 
induced processes in $p\bar{p}$ collision by convoluting
the hard part with only the parton densities, or in other 
words, by integrating out the common fragmentation probability 
$\int D_C^{\psi}$. Direct numerical calculation shows that 
the cross section induced by hard process (\ref{eq2}) overtakes
what induced by (\ref{eq1}) by a factor of 7 at $p_T = 15$ GeV.

For completeness, we consider processes
\bea
q_i(\bar{q_i}) + C(\bar{C}) \rightarrow q _i(\bar{q_i})
+\; C(\bar{C})\;,
\label{eq5}
\eea
in our numerical calculation as well. Here, the 
$q_i(\bar{q_i})$ represents partons of both valence 
and sea quarks(antiquarks) of the colliding nucleons. 
Practical exercise, similar as performed in preceding 
paragraph, shows that this kind of hard interaction 
processes also contributes more to $J/\psi$ 
large-$p_T$ production than via subprocess 
$g + g \rightarrow C\;+ \bar{C}$. And, to be noted that 
the latter was taken to be the dominant process in many of 
previous analyses within CSM. 

In the numerical calculation of the differential cross 
section, we need to choose a set of parton distributions. 
In this work we take CTEQ5M \cite{cteq} parameterization 
as our input. We have also tried another set of parton 
distributions, the MRST99 \cite{mrst}, and found that 
different parton distribution functions give similar 
results within tens of percent. That means the conclusions 
given in this paper will not be spoiled by taking a 
different set of parton distributions for convolution
in Eq.(\ref{ev}).

\begin{figure}[tbh]
\begin{center}
\psfig{file=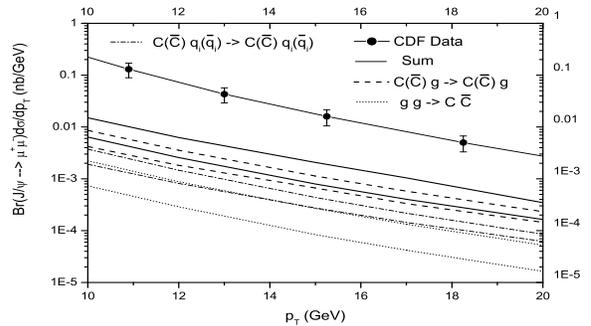,width=8.5cm,height=5cm,clip=0}
\end{center}
\caption[bt]{Fragmentation Contributions to the 
differential cross section for direct $J/\psi$ production 
at the Fermilab Tevatron, compared with the CDF experiment 
data read from \cite{cdf3}. The upper curve corresponds to 
$\mu_R = \mu_F = \mu_{\rm frag} = p_T/2$, 
while the lower curve to 
$\mu_R = \mu_F= \mu_{\rm frag}  = 2 p_T$.}
\label{graph2}
\end{figure}    

In figure 1, various fragmentation contributions 
to direct $\jp$ production are shown, and the sum of
them is confronted to the CDF data. Since we merely 
want to sketch the importance of sea-quark contributions 
to the large-$p_T$ charmonium production, the lowest order 
hard scattering cross sections and fragmentation function 
are employed. We also neglect the effect from non-diagonal 
splitting function, $P_{cg}$, in Eq. (\ref{ev}), which, 
as pointed out in Ref. \cite{cacciari,braaten2}, may give a 
factor of as large as 1.5 to the charm fragmentation process. 
We estimate the uncertainties of higher order
corrections by varying the scale, 
given that a lower value of scale will account for, 
in a certain degree, the contributions from higher order 
corrections and non-diagonal splitting functions. 
The upper and lower lines in the figure are 
obtained by varying the scales of factorization, 
renormalization and fragmentation. The upper 
curve corresponds to $\mu_R = \mu_F = \mu_{\rm frag} = 
p_T/2$, while the lower curve to $\mu_R = \mu_F
= \mu_{\rm frag}  = 2 p_T$. In drawing the diagram we 
use the fragmentation function given in Ref.\cite{braaten1}
and values quoted thereof ($R_0^2 = 0.8 {\rm GeV}^3$, 
$\alpha_s = 0.26$, $m_c = 1.5$ GeV). The symmetry 
of a sea quark and its antiquark in hard scattering
and fragmentation precesses is invoked. In addition, we 
perform our calculation in the $p_T \ge 10$ GeV region, 
where contributions from lowest order parton fusion processes
can be safely neglected,

From figure 1 we see that the discrepancy between direct
$\jp$ production data and the CSM prediction(the sum of
different processes) is less than an order in optimal case
(the upper solid line), rather than the common belief of
30 or more. The finding that the charm-sea 
initiated processes contribute dominantly to high 
$p_T$ charmonium production, as shown in figure 1, within CSM 
looks surprise, whereas, it is not really an unthinkable 
thing. Similar cases exist in some other quarkonium high 
energy production processes as well. For example, in 
photon-photon collision, quarkonium production via resolved 
processes is not always minor to the direct one. In addition, 
it should be noted that although the charm sea originates from 
the high order valence quark and gluon interactions, the naive 
NLO QCD result for $J/\psi(\psi')$ hadroproduction can not 
simply substitutes the result from sea-qaurk initiated  
processes. In the latter case we are considering 
the production via fragmentation mechanism, which is already
beyond LO in $\alpha_s$.

To conclude, in this work we study the relative importance 
of the sea quark initiated processes with respect 
to the valence quark and gluon initiated ones for 
large-$p_T$ $\jp (\psi')$ production within the CSM. 
It is found that the former may contribute more than
the latter by a factor of six. We notice that to many people
within the community the gluon initiated processes
are still taken to be the dominant ones for the high-$p_T$ 
$\jp (\psi')$ production from CSM calculations. We hope this 
work may elucidate it somehow. With including the new production 
scheme, the total cross section from CSM prediction fall off 
the experiment data by less than an order in extreme situation. 
Therefore, to explain the CDF data, COM is still necessary and 
essential. Nevertheless, the increase of the contribution from CSM 
means a shrinkage of the contribution from COM, which might be 
as large as twenty percent. Furthermore, the result in this 
work gives us a strong hint that the $J/\psi(\psi')$-surplus
production at the Fermilab Tevatron might still be explained 
within CSM after including the NLO calculation, like the charmonium
photoproduction at HERA \cite{kramer3}\cite{kramer4}. At least
the color-octet contribution will be much less than previously
thought. Finally, we would like to point 
out that since the (anti)charm quark is taken to be massless 
in our calculation, some uncertainties will be induced by 
this measure to order of ${\cal O}(M_{J/\psi}/p_T)$. For detailed
estimations of the uncertainties by taking zero mass scheme see
Refs. \cite{kramer2}\cite{demina}. In all, the error would be 
within a factor of two in our analysis, which will not change the 
conclusions of this work.

This work was supported by the Grant-in-Aid of JSPS 
committee. The author would like to thank  
K. Hagiwara for helpful discussion.

\end{multicols}
\end{document}